\documentclass{edm_article}
\begin{document}

\title{What metrics of participation balance predict outcomes of collaborative learning with a robot?}

\numberofauthors{9}
\author{
\alignauthor Yuya Asano\\
\affaddr{University of Pittsburgh}\\
\email{yua17@pitt.edu}
\alignauthor Diane Litman\\
\affaddr{University of Pittsburgh}\\
\email{dlitman@pitt.edu}
\alignauthor Quentin King-Shepard
\affaddr{University of Pittsburgh}\\
\email{quk1@pitt.edu}
\and
\alignauthor Tristan Maidment\\
\affaddr{University of Pittsburgh}\\
\email{Tristan.maidment@pitt.edu}
\alignauthor Tyree Langley\\
\affaddr{University of Pittsburgh}\\
\email{tyl27@pitt.edu}
\alignauthor Teresa Davison\\
\affaddr{University of Pittsburgh}\\
\email{tid30@pitt.edu}
}
\additionalauthors{Additional authors: Timothy Nokes-Malach (University of Pittsburgh,
email: {\texttt{nokes@pitt.edu}}), Adriana Kovashka (University of Pittsburgh, email: {\texttt{kovashka@cs.pitt.edu}} and Erin Walker
(University of Pittsburgh, email: {\texttt{eawalker@pitt.edu}}).}

\maketitle

\begin{abstract}
One of the keys to the success of collaborative learning is balanced participation by all learners, but this does not always happen naturally. Pedagogical robots have the potential to facilitate balance. However, it remains unclear what participation balance robots should aim at; various metrics have been proposed, but it is still an open question whether we should balance human participation in human-human interactions (HHI) or human-robot interactions (HRI) and whether we should consider robots' participation in collaborative learning involving multiple humans and a robot. This paper examines collaborative learning between a pair of students and a teachable robot that acts as a peer tutee to answer the aforementioned question. Through an exploratory study, we hypothesize which balance metrics in the literature and which portions of dialogues (including vs. excluding robots' participation and human participation in HHI vs. HRI) will better predict learning as a group.
We test the hypotheses with another study and replicate them with automatically obtained units of participation to simulate the information available to robots when they adaptively fix imbalances in real-time. Finally, we discuss recommendations on which metrics learning science researchers should choose when trying to understand how to facilitate collaboration.
\end{abstract}

\keywords{Participation balance, Collaborative learning, Human-robot interaction, Teachable robot} 

\section{Introduction}
One important factor contributing to the success of collaborative learning among humans is balanced participation, which means every learner contributes evenly \cite{bruno2010social,najafi2010analyzing}. Technology may be able to help facilitate balanced participation as it gets more intelligent. Indeed, there is evidence that robots can promote balance with their behavior (e.g., \cite{skantze2017predicting,tennent2019micbot}), and Al Moubayed and G. Skantz \cite{al2011turn} have found embodied agents are more effective in moderating turn-taking than virtual agents. 
However, there is no consensus on metrics of participation balance in complicated learning environments that involve multiple humans and a robot that actively participates. The lack of such consensus lies in the following three dimensions:\\ 
\textbf{(1) What counts as participation?} A simple method is to count the number of turns learners took or words they spoke. With audio data, we can also measure how long they talked. However, prior literature has not explicitly compared which units of participation are advantageous.\\ 
\textbf{(2) How should we convert these units of participation into a single measure of balance per group?} Several metrics of balance have been proposed in a wide range of fields, but there is no consensus on which one works the best in a given situation (see Section \ref{metrics} for more details). \\
\textbf{(3) Which portion of the dialogue should be used?} Collaborative dialogues with a robot can be divided into human-robot portions (interactions between humans and a robot) and human-human portions (interactions among humans). Since a robot tends to take a different role from humans (often a facilitator or, in this work, a teachable robot), humans talk to a robot differently from other humans \cite{chandra2015can}. Therefore, balance in human-robot portions of talk within a group of multiple humans and robots may have different implications on learning from balance in human-human portions. For example, promoting balance in human-robot portions may not be as effective as in human-human portions.

In this paper, we compare the measures of participation balance used in the literature from the three questions above to provide guidance for researchers on which metrics they should use in collaborative learning with a robot. First, we discuss the approaches to the first two questions in the literature and their hypothetical advantages to others. Then, we analyze data from an exploratory study to develop hypotheses about the three questions. We test them with another study using the units of participation automatically collected from an automatic speech recognition system and raw audio data. This test opens up new opportunities for a robot to put the theory into practice via adaptive interventions to fix imbalances in real time. To summarize, our contributions are
\setlist{nolistsep}
\begin{enumerate}[noitemsep,topsep=0pt]
    \item Comprehensive comparisons of the metrics of participation balance.
    \item Theoretical understanding of the relationships between participation balance and learning of student groups.
    \item A practical guidance on which metrics to use in collaborative learning with a teachable robot.
\end{enumerate}

\section{Related work}
\subsection{Participation balance in collaborative learning}

Both objective and subjective participation balance in collaborative learning is associated with a higher-quality group product, a better learning gain, and a more satisfying experience. For instance, in collaborative Wikipedia editing, groups of student editors that exhibit balanced participation engage in collaborative knowledge construction and have a learning gain in the contents of the Wikipedia pages \cite{bruno2010social,najafi2010analyzing}. Balanced participation in online group discussions results in higher achievement in group assignments \cite{jahng2010collaborative}, epistemically more complex and scientifically more sophisticated theory development by learners \cite{zhu2017evaluating}, and higher learners' satisfaction with the overall collaboration \cite{strauss2021promoting}. Moreover, students' subjective evaluation of the balance of participation in discussion is positively correlated with scores of their individually written learning journals \cite{lindblom2003makes}.
In addition to the correlational studies above, there is some work that implies balanced participation causes enhanced learning \cite{cortazar2022impacts,njenga2018facilitating}. 

Although the literature highlights the importance of balanced participation, it has different definitions (cf. Section \ref{metrics}), making it difficult to conclude what imbalance should be addressed for increased learning outcomes as a group. This paper evaluates those different definitions comprehensively to clarify what to balance. In addition, this paper expands the research on participation balance in collaborative learning to a group of humans and \textit{robots}.

\subsection{Balancing participation with a robot}\label{mediating}

The previous section demonstrates that participation balance is key to successful collaborative learning. However, it is not always achieved naturally. Several collaborative learning support technologies have focused on balancing participation, for example, by visualizing individuals' contributions to the group \cite{bachour2010interactive, tausch2014groupgarden}, or by prompting individuals who are not participating to contribute \cite{nabetani2021introducing, do2022should}. We are particularly interested in how a robot can mediate participation in a group conversation, as its verbal and non-verbal behaviors might lend themselves to more engaging and less disruptive facilitation of collaborative interactions than explicit visualization or prompting approaches. For example, a robot can engage non-participating individuals by gazing at them \cite{gillet2021robot}, asking a directed question to them \cite{skantze2017predicting}, giving them a fact or trivia relevant to the topics of the conversation \cite{ayllon2021identification}, moving a mic to them \cite{tennent2019micbot}, eliciting an action toward inclusion \cite{gillet2020social}, and making vulnerable statements (self-disclosure, storytelling, and humor) \cite{traeger2020vulnerable}. Tennent et al. \cite{tennent2019micbot} further have found that intervention by a robot to facilitate balance leads to better team problem-solving performance.

Nevertheless, these robots typically acted as facilitators and did not actively engage in discussions, unlike our teachable robot. Therefore, the work above did not discuss participation balance \textit{including the robot's participation} or \textit{differentiate participation directed to humans from participation directed to robots}. Our work investigates whether we should consider the robot's participation for balance and how differently the balance in the human-robot portion and that in the human-human portion is related to learning outcomes.

\subsection{Metrics of balance}\label{metrics}
We review different metrics of participation balance in the related work above and beyond. We discuss their differences in the units of participation and conversion to balance. 

\subsubsection{Units of participation}
The literature is divided on what counts as participation. The commonly used units are the number of turns (e.g., \cite{ayllon2021identification,njenga2018facilitating,tennent2019micbot}), the number of words (e.g., \cite{bruno2010social,jahng2010collaborative,strauss2021promoting}), and the length of speech in seconds (e.g., \cite{ayllon2021identification,gillet2021robot,skantze2017predicting,tennent2019micbot,traeger2020vulnerable}). We noticed that much collaborative learning literature \cite{bruno2010social,jahng2010collaborative,njenga2018facilitating,strauss2021promoting} uses only the number of turns and words whereas HRI literature \cite{ayllon2021identification,gillet2021robot,skantze2017predicting,tennent2019micbot,traeger2020vulnerable} tends to use speech length, and utilizes voice activity detection to alleviate manual annotations \cite{ayllon2021identification,gillet2021robot,skantze2017predicting}. The HRI literature that does analyze the number of turns involves coding recordings manually \cite{tennent2019micbot} or using a microphone array to perform automatic speaker localization \cite{ayllon2021identification}.

\subsubsection{Conversion to balance}

Since the units of participation are measured per speaker, we have to turn them into a single number per group. We only include formulas that can be used for any units of participation above to evaluate them based on different units. 

A popular way to get a group-level balance measure is to sum up deviations. The most well-known formula is the standard deviation (SD) (e.g., \cite{jahng2010collaborative}). Jarvenpaa et al. \cite{jarvenpaa1988computer} proposed to use the coefficient of variation (CV), which divides SD by the average participation and is used in later studies (e.g., \cite{gillet2020social,kershisnik2016collaboration}). Some HRI literature (e.g., \cite{gillet2021robot,tennent2019micbot}) uses an absolute deviation (AD) (cf. Table \ref{formula}). When the number of speakers is 2, $AD = 2SD$, but this linear relationship does not hold for other cases.

Another family of balance formulas is from economics. The Gini coefficient \cite{gini1912variabilita} captures income and wealth inequality but is also used as a measure of participation balance (e.g. \cite{strauss2021promoting}). 
Ray and Singer \cite{ray1973measuring} pointed out that its upper-bound, $1 - \frac{1}{n}$ ($n$ is the number of speakers in a group), is overly sensitive to $n$ for a small $n$ and problematic when quantifying the degree of monopoly. They proposed an index of concentration (CON), which standardizes CV to take a value between 0 and 1 (i.e., $\frac{CV}{\sqrt{n-1}}$, cf. Table \ref{formula}) \cite{taagepera1977generalized}.
On the contrary, Hiltz et al. \cite{hiltz1989experiments} addressed the problem with the Gini coefficient by multiplying it by $\frac{n}{n-1}$ (i.e., an unbiased estimator of the Gini coefficient \cite{mejias2018observation}) to quantify inequality of participation (IP) (cf. Table \ref{formula}). This is also called adjusted Gini \cite{deltas2003small}.
Note that CON and IP are equal when $n=2$.

Bruno \cite{bruno2010social} formulated balance as Shannon entropy \cite{shannon1948mathematical} (cf. Table \ref{formula}) from information theory. The range of entropy depends on the base of $\log$ and can be set to $[0,1]$ by choosing $n$ as the base. 

Each conversion formula has advantages and disadvantages. 
SD is easy to compute as most statistics tools support it, but SD and AD are susceptible to the scale of the units of participation. CV and the Gini coefficient remove this limitation by having functions of the average participation in their denominators. However, their ranges still depend on $n$. CON and IP standardize CV and the Gini coefficient, respectively to take values between 0 and 1. Indeed, Rose et al. \cite{rose2020inequality} showed CON is a less biased participation balance measure in small groups than the Gini coefficient.
Entropy penalizes dominance by a few speakers less than CON and IP. When two speakers spoke 49.9\% of the time each and one speaker spoke .2\%, $CON=IP=.497$ and $entropy = .643$ (the base of $\log$ is 3). When one speaker spoke 77\% and two speakers spoke 11.5\% each, $CON=IP=.655$ and $entropy = .636$. Entropy stays almost the same while CON and IP increase by .16. On the flip side, entropy punishes the presence of one quiet speaker more than CON and IP. Assume three speakers spoke 58\%, 30\%, and 12\%. Then, $CON=.401$, $IP=.460$, and $entropy=.848$. Compared to the first case, CON and IP drop at most by .1, but entropy changes more than .2.

\begin{table}[t]
\begin{minipage}{\linewidth}
\centering
\caption{Formulas to convert units of participation to balance and their values under the perfect balance and imbalance. $n$ is the number of speakers in a group, $x_i$ is speaker $i$'s participation, $\mu$ is the average participation, $p_i$ is the proportion of participation by speaker $i$, $E_i$ is the expected cumulative proportion of participation under the perfect balance, and $O_i$ is the observed cumulative proportion.}
\label{formula}
\begin{tabular}{p{2.5cm}C{2cm}C{1.3cm}C{1.5cm}} \hline
Name & Formula & Perfect Balance & Perfect Imbalance \\ \hline
Standard deviation (SD) & $\displaystyle \sqrt{\frac{\Sigma (x_i - \mu)^2}{n}}$ & 0 & Varies\footnote{The maximum of SD is 71\% of range for two data points and less than 58\% for three or more data points \cite{petocz2005upper}.\label{foot:sd}} \\
Absolute deviation (AD) & $\displaystyle \sum\limits |x_i - \mu|$ & 0 & Varies \\ 
Index of concentration (CON) & $\displaystyle \sqrt{\frac{\Sigma p_i^2 - \frac{1}{n}}{1 - \frac{1}{n}}} $ & 0 & 1\\
Inequality of participation (IP) & $\displaystyle \frac{\frac{1}{n} \sum(E_i - O_i)}{\frac{1}{2}(1 - \frac{1}{n})}$ & 0 & 1 \\
Entropy-based\footnote{The original formula of entropy is $-\Sigma p_i\log_n{p_i}$, but we changed it so that the direction of the measure becomes the same as other formulas in this table.} & $\displaystyle 1+\Sigma p_i\log_n{p_i}$ & 0 & 1\\
\hline
\end{tabular}
\end{minipage}
\end{table}

\section{Study 1: Comparing measures of participation balance}
This exploratory study evaluates the ability of different measures of participation balance to predict group learning outcomes from three aspects: the units of participation, the formulas to convert the units to balance, and the portions of dialogue to balance participation.

\subsection{Corpus}

We used data from our previous study \cite{steele2022it}. 
It has 40 undergraduate students from a mid-Atlantic US university teaching middle-school level ratio word problems to a robot named Emma. They consisted of 35 females and 5 males, and their mean age was $19.64$ ($SD = 1.25$). The students identified as 17 White, 13 Asian, 5 Black, 1 Latino, and 4 no answer. The study followed a procedure approved by an institutional review board (IRB), and all students consented to the use of their data for research purposes. 

\begin{figure}[t]
    \centering
    \includegraphics[width=0.45\linewidth]{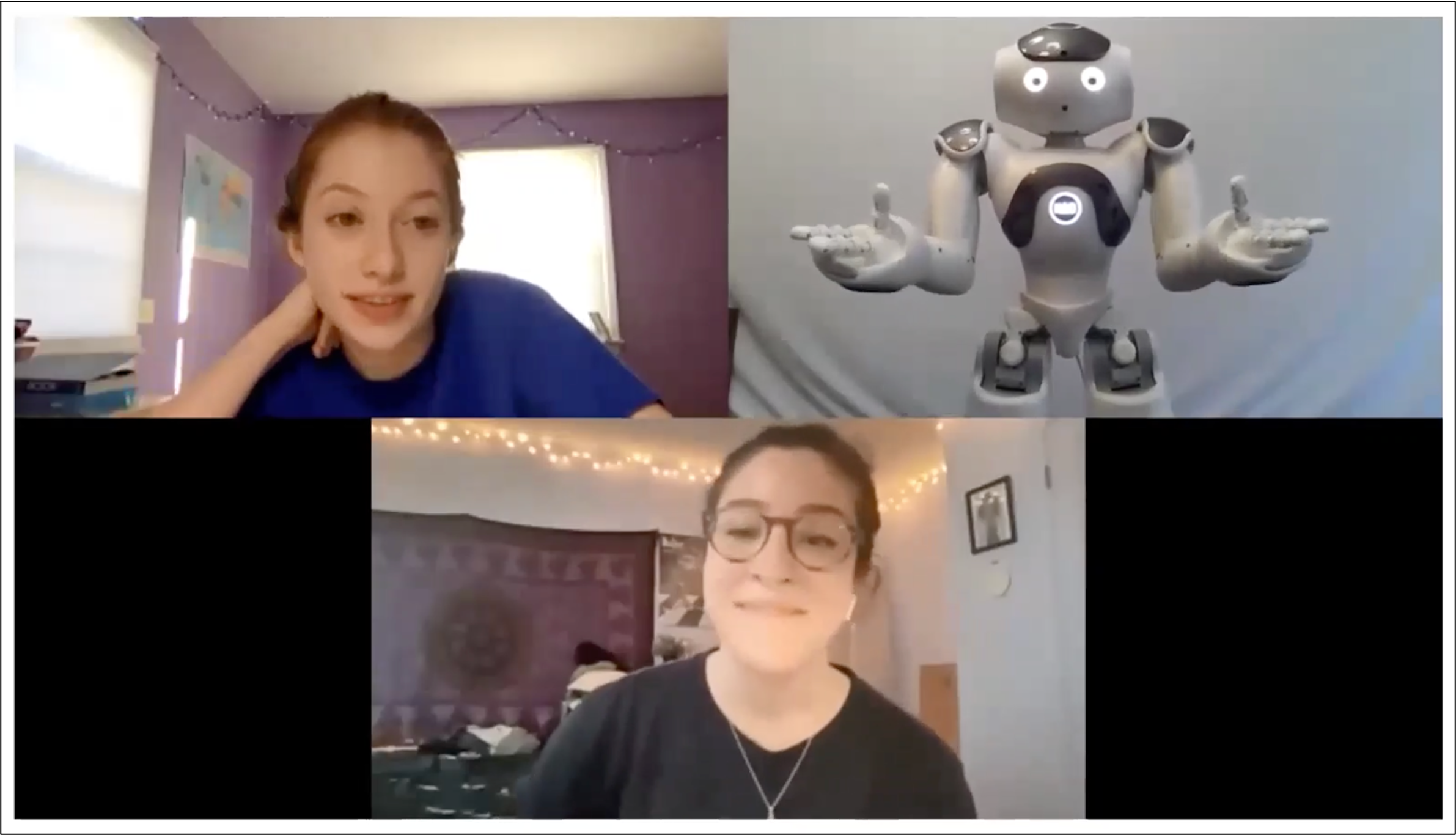}
    \includegraphics[width=0.45\linewidth]{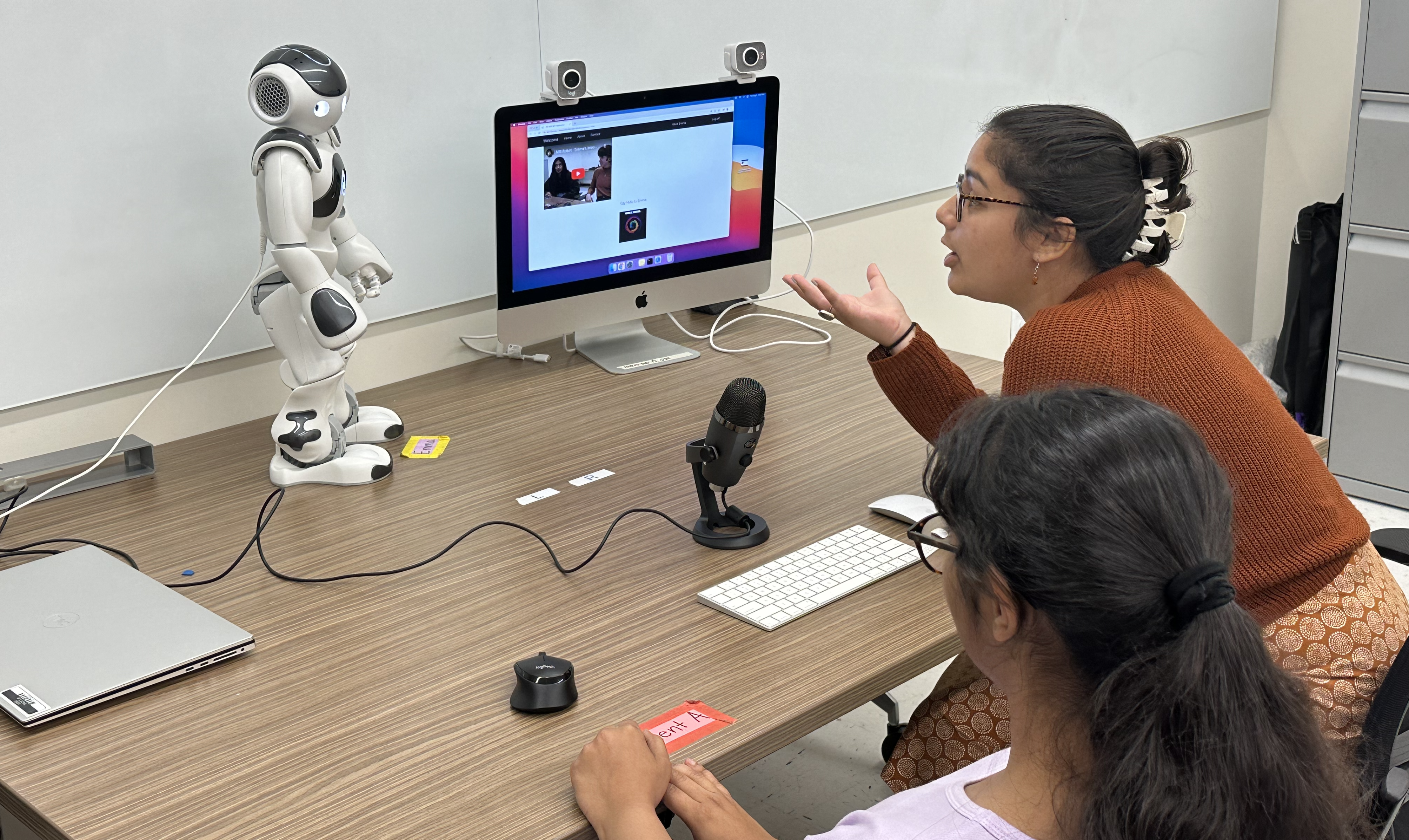}
    \caption{A pair of students teaching Emma over Zoom (left) \cite{asano2022comparison} and in person (right).}
    \label{fig:emma}
\end{figure}

The dataset has 28 students teaching in pairs and 12 students teaching alone. For the purpose of this paper, we excluded students who worked alone. 
Following Asano et al. \cite{asano2022comparison}, we dropped one pair because one of the students did not talk to Emma during the session. Each session was 30 minutes and held over Zoom due to COVID-19 (Figure \ref{fig:emma}). The students taught the problems to Emma and interacted with her (Emma) through spoken dialogue. She recognized their voice only while they were holding a button on a web application. Her dialogue system was implemented in Artificial Intelligence Markup Language, which did pattern matching with students' utterances to select a pre-authored response \cite{asano2022comparison}. 
The pre-authored responses included questions to students and self-explanations.\footnote{This version of Emma was designed for individual use and did not explicitly facilitate collaboration between students.} Each problem required multiple steps to be solved, and the pre-authored responses reflected this nature so that the students would teach Emma step-by-step with multiple turns in one problem, preventing reaching the solution in one turn. A real example interaction between students and Emma is illustrated in Table \ref{tab:example}.

\begin{table}[t]
    \centering
    \caption{Example interactions in a session. The addressee of Emma's last utterance is A because A replied in the following turn that is not in this table.}
    \label{tab:example}
    \begin{tabular}{R{1cm}|L{5.7cm}|L{1cm}}
        Speaker & Utterance & Addressee \\
        \hline
        Emma & Can you give me a hint for step two of this third problem?  & A \\
        \hline
        A & So then I would just say... What did I do? Oh my God. I'm having a complete mind empty. Do you literally just...  & B \\
        \hline
        B & Probably ten. & A \\
        \hline
        A & You multiply the seven point two five by ten. Right? & B \\
        \hline
        B & Yeah.  & A \\
        \hline
        A & Yeah. Okay. So, do I just let to go? Yeah. I'll just save that word for... & B \\
        \hline
        A & Next, you're going to multiply seven point two five times ten. & Emma \\
        \hline
        Emma & Opposite of the last step, I multiply. But do you think you could help me understand. How do you know I multiply? & B \\
        \hline
        A & Since we know that you can ride ten miles every hour. & B \\
        \hline
        B & Mm-hmm. Yeah. & A \\
        \hline
        B & Emma, it's because you know that you can ride ten miles per hour. & Emma\\
        \hline
        Emma & I know I can ride my bike for ten miles in one hour. I multiply the ten miles per hour times seven point two five? & A
    \end{tabular}
\end{table}

The dataset contains manual transcripts, annotations of the addressees and the timestamps (start and end) of each utterance, pre-tests, and post-tests. The pre-tests and post-tests were taken individually. Both tests were isomorphic and counterbalanced. We excluded the utterances where participants were talking to the researcher due to technical issues, the same as Asano et al. \cite{asano2022comparison}. 
We followed the same procedure as Steele et al. \cite{steele2022it} 
to discard ineffective problems in the tests and used the remaining eight problems to assess learning. The average scores in pre- and post-test were 5.58 ($SD = 1.90$) and 6.82 ($SD = 1.35$), respectively. In this paper, we defined group learning by partial correlations with post-test scores controlled by pre-test scores averaged within pairs.

\subsection{Metrics of participation balance} \label{measures}
We compare the metrics of participation balance in collaborative learning dialogue from three aspects: the units of participation, the ways to convert them to one balance value per group, and the portions of dialogue to be used.

\textbf{Units of participation:} We use the number of turns and words and the speech length as discussed in the previous section.

\textbf{Conversion to balance:} The formulas we use are summarized in Table \ref{formula}. We do not consider CV or the Gini coefficient because they are the linear transformation of CON and IP, respectively (thus giving the same correlation coefficients), and are not scaled between 0 and 1, unlike CON and IP.
For entropy, we choose $n$ as the base of $\log$ to set the maximum to 1, the same as \cite{zhu2017evaluating}, and modified to $1 - entropy$ so that it measures imbalance like other measures instead of balance.

\textbf{Portions of dialogue:} Our study has two main components: teaching Emma (a robot) and discussing math problems with a partner. 
We consider balance in four portions: the whole dialogue with Emma's turns, the whole dialogue without Emma's turns, the Emma-student dialogue, and the student-student dialogue.
The following is an example of the calculation of participation balance for each portion with the example interaction in Table \ref{tab:example}. The whole dialogue with Emma's turns uses all utterances in the table. Emma has 3 turns, A has 5 turns, and B has 4 turns. Thus, $x_{Emma} = 3$, $x_A = 5$, $x_B = 4$, $p_{Emma} = \frac{1}{4}$, $p_A = \frac{5}{12}$, and $p_B = \frac{1}{3}$, $E_i=\frac{i}{3} \forall i \in \{1,2,3\}$, $O_1 = \frac{1}{4}$, $O_2 = \frac{1}{4} + \frac{1}{3} = \frac{7}{12}$, and $O_3 = 1$. The whole dialogue without Emma's turns excludes only Emma's utterances. Thus, $x_A = 5$, $x_B = 4$, $p_A = \frac{5}{9}$, and $p_B = \frac{4}{9}$. The Emma-student dialogues consist of Emma's turns and the students' turns directed to Emma. For instance, the Emma-studentA portion focuses on the turns whose speakers and addressees are Emma and A, so $x_{Emma} = 2$, $x_A = 1$, $p_{Emma} = \frac{2}{3}$, and $p_A = \frac{1}{3}$ in the Emma-student portion. Finally, the student-student dialogues are the whole dialogues excluding Emma’s turns and students’ utterances addressed to her. Since A talked to B 4 times and B talked to A 3 times, $x_A = 4$, $x_B = 3$, $p_A = \frac{4}{7}$, and $p_B = \frac{3}{7}$ in the student-student portion.

\subsection{Results}\label{result1}

\begin{table}[t]
    \centering
    \caption{The mean and SD (in parentheses) of the units of participation in Study 1. Turns and words are raw counts, and speech was measured in seconds.}
    \label{tab:descriptive-unit1}
    \begin{tabular}{p{0.8cm}c|ccc}
         & & Turn & Word & Speech \\
        \hline
        \multirow{4}{1.4cm}{Whole} & \multirow{2}{*}{Emma} & 63.0 & 1193.0 & 331.2 \\
        & & (13.9) & (222.2) & (60.9) \\
        \cline{2-5}
        & \multirow{2}{*}{Student} & 85.5 & 964.3 & 386.3 \\
        & & (27.0) & (339.3) & (134.5) \\
        \hline
        \multicolumn{2}{r|}{\multirow{2}{*}{Emma-student}} & 27.9 & 389.5 & 154.5 \\
        & & (9.76) & (202.9) & (71.5) \\
        \hline
        \multicolumn{2}{r|}{\multirow{2}{*}{Student-student}} & 57.6 & 574.9 & 231.8 \\
        & & (29.7) & (373.4) & (154.1) \\
        \hline
    \end{tabular}
\end{table}

\begin{table*}[t]
    \caption{The mean and SD (in parentheses) of the balance of participation measures in Study 1.}
    \label{tab:descriptive-formula1}
    \begin{minipage}{0.48\textwidth}
    \centering
    (a) The whole conversation including Emma's turns.
    \begin{tabular}{C{1.5cm}|c|c|c|c|c}
        Whole w/ Emma & SD & AD & CON & IP & Entropy \\
        \hline
        \multirow{2}{*}{Turn} & 18.4 & 50.1 & .160 & .175 & .029 \\
        & (10.4) & (29.2) & (.071) & (.076) & (.025) \\
        \multirow{2}{*}{Word} & 259.3 & 696.6 & .179 & .198 & .035 \\
        & (133.4) & (356.8) & (.089) & (.099) & (.029) \\
        \multirow{2}{*}{Speech} & 87.2 & 237.7 & .162 & .176 & .029 \\
        & (54.1) & (151.4) & (.083) & (.090) & (.029) \\
    \end{tabular}
    \end{minipage}
    \begin{minipage}{0.48\textwidth}
    \centering
    (b) The whole conversation excluding Emma's turns.
    \begin{tabular}{C{1.6cm}|c|c|c|c|c}
        Whole w/o Emma & SD & AD & CON & IP & Entropy \\
        \hline
        \multirow{2}{*}{Turn} & 7.39 & 14.8 & .099 & .099 & .013 \\
        & (5.88) & (11.8) & (.090) & (.090) & (.019) \\
        \multirow{2}{*}{Word} & 172.0 & 343.9 & .171 & .171 & .032 \\
        & (151.8) & (303.5) & (.125) & (.125) & (.041) \\
        \multirow{2}{*}{Speech} & 65.9 & 131.8 & .163 & .163 & .027 \\
        & (55.9) & (111.8) & (.107) & (.107) & (.030) \\
    \end{tabular}
    \end{minipage}

    \begin{minipage}{0.48\textwidth}
    \centering
    (c) The Emma-student conversation.
    \begin{tabular}{C{1.1cm}|c|c|c|c|c}
        Emma-student & SD & AD & CON & IP & Entropy \\
        \hline
        \multirow{2}{*}{Turn} & 5.19 & 10.4 & .176 & .176 & .036 \\
        & (4.96) & (9.92) & (.140) & (.140) & (.043) \\
        \multirow{2}{*}{Word} & 77.9 & 155.9 & .155 & .155 & .033 \\
        & (103.8) & (207.6) & (.151) & (.151) & (.057) \\
        \multirow{2}{*}{Speech} & 25.4 & 50.9 & .138 & .138 & .023 \\
        & (29.7) & (59.4) & (.115) & (.115) & (.030) \\
    \end{tabular}
    \end{minipage}
    \begin{minipage}{0.48\textwidth}
    \centering
    (d) The student-student conversation.
    \begin{tabular}{C{1.6cm}|c|c|c|c|c}
        Student-student & SD & AD & CON & IP & Entropy \\
        \hline
        \multirow{2}{*}{Turn} & 2.96 & 5.92 & .065 & .065 & .004 \\
        & (1.44) & (2.87) & (.033) & (.033) & (.003) \\
        \multirow{2}{*}{Word} & 111.3 & 222.7 & .164 & .164 & .027 \\
        & (118.5) & (236.9) & (.106) & (.106) & (.031) \\
        \multirow{2}{*}{Speech} & 49.2& 98.3 & .192 & .192 & .035 \\
        & (53.5) & (107.0) & (.109) & (.109) & (.034) \\
    \end{tabular}
    \end{minipage}
    
\end{table*}

\begin{table*}[t]
    \caption{Partial correlation between participation balance and post-test scores controlled by pre-test scores in Study 1 ($n=13$). Correlations marked with * and ** have $p<.05$ and $p<.01$, respectively.}
    \label{tab:results1}
    \begin{minipage}{0.48\textwidth}
    \centering
    \label{tab:whole-w-emma}
    (a) The whole conversation including Emma's turns.
    \begin{tabular}{C{1.6cm}|c|c|c|c|c}
        Whole w/ Emma & SD & AD & CON & IP & Entropy \\
        \hline
        Turn & -.159 & -.174 & -.099 & -.073 & -.150 \\
        Word & -.014 & .003 & .151 & .135 & .095 \\
        Speech & -.443 & -.430 & -.388 & -.406 & -.345 \\
    \end{tabular}
    \end{minipage}
    \begin{minipage}{0.48\textwidth}
    \centering
    (b) The whole conversation excluding Emma's turns.
    \begin{tabular}{C{1.6cm}|c|c|c|c|c}
        Whole w/o Emma & SD & AD & CON & IP & Entropy \\
        \hline
        Turn & .432 & .432 & .432 & .432 & .387 \\
        Word & -.605* & -.605* & -.501 & -.501 & -.553 \\
        Speech & -.602* & -.602* & -.500 & -.500 & -.449\\
    \end{tabular}
    \label{tab:whole-wo-emma}
    \end{minipage}

    \begin{minipage}{0.48\textwidth}
    \centering
    (c) The Emma-student conversation.
    \begin{tabular}{C{1.6cm}|c|c|c|c|c}
        Emma-student & SD & AD & CON & IP & Entropy \\
        \hline
        Turn & .406 & .406 & .373 & .373 & .412 \\
        Word & -.206 & -.206 & -.039 & -.039 & -.310 \\
        Speech & -.109 & -.109 & .045 & .045 & -.099 \\
    \end{tabular}
    \label{tab:emma-student}
    \end{minipage}
    \begin{minipage}{0.48\textwidth}
    \centering
    (d) The student-student conversation.
    \begin{tabular}{C{1.6cm}|c|c|c|c|c}
        Student-student & SD & AD & CON & IP & Entropy \\
        \hline
        Turn & .247 & .247 & .487 & .487 & .410 \\
        Word & -.572 & -.572 & -.738** & -.738** & -.706* \\
        Speech & -.483 & -.483 & -.380 & -.380 & -.304 \\
    \end{tabular}
    \label{tab:student-student}
    \end{minipage}
    
\end{table*}

Tables \ref{tab:descriptive-unit1} and \ref{tab:descriptive-formula1} show the descriptive statistics of the units of participation and balance metrics in each portion of the dialogue, respectively. Table \ref{tab:results1} summarizes the partial correlation between the measures of participation balance and the groups' post-test scores controlled by their pre-test scores. Since all formulas measure imbalance, a negative correlation in the tables means a positive correlation between balance and learning.

\paragraph{Units of participation} We compare the rows of each sub-table in Table \ref{tab:results1} to see the differences between the three units of participation. 
We found significant correlations only when the units of participation were word counts or speech length in Tables \ref{tab:results1}(b) and \ref{tab:results1}(d). We did not see any significant correlations for speech length in Table \ref{tab:results1}(d) possibly because the annotation did not reflect pauses within a turn. Students often paused when they were unsure in student-student conversations. However, as long as they talked to the same speaker, we only timestamped the beginning and the end of the turn without subtracting the pause. This may have been mitigated by including speech to Emma where students were more confident and could be potentially resolved for automated units by removing silence in audio data.

In our setting, turns may not be an ideal unit for two reasons. First, they do not fully capture the ``amount of contribution'' by each student because both a short utterance that just replies ``Okay'' and a long utterance consisting of rigorous explanations are considered equally as one turn. The latter should have more ``amount of contribution,'' and the word count and the speech length reflect the amount better. Second, the number of turns is almost balanced in student discussions since our experiment has only two students. The difference in the number of turns only comes from when students talk to Emma or when they go to the next problem. Similarly to the example in the first reason, the number of words and the length of speech are likely to have more imbalance in an unsuccessful collaboration.

\paragraph{Conversion to balance} We compare different formulas to convert the units to balance by comparing the columns within each sub-table. The correlations had the same signs unless their strengths were weak. SD and AD had stronger correlations in the whole conversations without Emma's turns (Table \ref{tab:results1}(b)) while CON, IP, and entropy had stronger ones in the student-student conversations (Table \ref{tab:results1}(d)).

It should be noted that it is likely the mechanism relating balance and learning is different from any mechanism relating total participation to learning.  We do see the total participation was marginally correlated with the post-test scores controlled by the pre-test scores in the whole dialogue without Emma's turns ($r=-.510$, $p=.09$ for words and $r=-.560$, $p=.06$ for speech) but not in the student-student conversation ($r=-.332$, $p>.1$ for words and $r=-.407$, $p>.1$ for speech). However, it was not significantly correlated with any of the balance formulas except for SD and AD in the student-student conversation ($r=.790$, $p<.01$ for words and $r=.695$, $p<.01$ for speech length).

\paragraph{Portions of dialogue} We compare the sub-tables while fixing the rows and columns. Significant correlations appear only after removing Emma's turns (compare Tables \ref{tab:results1}(a) and \ref{tab:results1}(b)). To break this down, Tables \ref{tab:results1}(c) and \ref{tab:results1}(d) indicate that the balance in students' discussion is more associated with their learning than the balance in their chances to talk to Emma. The fact that we have significant correlations in student-student conversations but not in Emma-student conversations implies that learning happens during discussions between students instead of teaching Emma.

\subsection{Summary}
This study revealed that strong, significant correlations were in the whole conversations without Emma's turns and the student-student conversations. The balance in terms of the number of words and the length of speech was more predictive of learning than the number of turns. All balance formulas exhibited strong correlations for these units and portions though SD and AD were significant only in the whole conversations without Emma's turns whereas CON, IP, and entropy were significant only in the student-student conversations.

\section{Study 2: Testing hypotheses from study 1}
We develop hypotheses according to Study 1 and test them with a new study. We also examine how robust they are, by changing the source of the units of participation from the human transcripts to automatically generated transcripts.

\subsection{Corpus}

We recruited 28 undergraduate students from a mid-Atlantic US university, none of whom participated in Study 1. 26 students worked in pairs with another student. Two students whose partners did not arrive for the study collaborated with a research team member. We removed these two students from our analysis. The participants consisted of 9 males, 18 females, and 1 non-binary person with a mean age of 20.67 (SD = 2.38). They identified themselves as 13 Asian, 2 Black, 2 Nigerian, 10 White, and 1 Multiracial. All following testing procedures were approved by the IRB and all students who participated gave consent for the use of their data. 

We used the same robot Emma but made some adjustments to Study 1. First, Study 2 was conducted in person as shown in Figure \ref{fig:emma}. Each student had their own mouse pointing to the same cursor on a shared monitor next to Emma to navigate through the same web application as Study 1. Second, we changed the order of problems to teach Emma so that the difficulty of the problems goes up gradually. Third, we extended the time of the sessions to 45 minutes to allow students to teach more problems. Finally, we removed the problems in pre- and post-tests that were removed from analysis in Study 1 and modified the remaining ones for a better alignment with the problems students teach to Emma.

Moreover, we made Emma gaze and reference students randomly to later test her ability to promote balanced participation (will be reported in future papers). She referenced ``Student A'' or ``Student B'' 25\% of the time, respectively, when she asked a question. She also looks at the student she referenced. Other times, she looked at Student A and B at 35\% each, the monitor at 15\%, and the top left (acting as if she was thinking) at 15\% of time for $s \sim N(0,20)$ seconds.

This time, we recorded separate audio channels for different students by having them wear a head microphone. The separate audio channels allowed us to get speech lengths and automatic transcription for each student. We removed the silence by applying a noise gate and the time they talked to Emma based on the logged data when needed. We counted the number of words on the transcripts generated automatically by Whisper medium.en \cite{radford2023robust}. 

\subsection{Hypotheses}
Study 1 revealed that word counts and speech length had significant correlations (i.e., correlations marked with * or ** in Table \ref{tab:results1}) in the portions that do not contain Emma's turns. Therefore, our first hypothesis is:\\
\textit{H1: The balance in word counts and speech length in the whole dialogue without Emma's turns and in the student-student portion is correlated with learning as a group.}

In those portions, all formulas yielded correlations stronger than $-.3$ though the strengths varied across the portions and the units. Thus, we consider all formulas in our second hypothesis:\\
\textit{H2: All formulas to convert the number of words to balance measures can capture the correlation with learning.}\\
Regarding H2, we only use SD, IP, and entropy as the formulas since AD and CON are the linear transformations of SD and IP, respectively, when $n=2$.

\subsection{Results}
The descriptive statistics of the number of words and speech length are in Table \ref{tab:descriptive-unit2}. Compared to Study 1, the number of words roughly doubled, and the speech length more than tripled. We posit these increases are not solely due to the longer duration because the rate of the increase surpasses the change in duration. For example, the in-person setting of Study 2 possibly made the student-student conversations easier than Study 1 conducted over Zoom. The descriptive statistics of the balance measures are in Table \ref{tab:descriptive-formula2}. Overall, the imbalance increased from Study 1 except for IP and entropy of word count in the whole conversation without Emma.

The correlation results are in Table \ref{tab:results2}. We adjusted p-values with the Holm method to mitigate type-I error in this confirmatory study. Both H1 and H2 are partially validated; only IP and entropy of the number of words in the student-student dialogues had significant correlations.

\begin{table}[t]
    \centering
    \caption{The descriptive statistics of the number of words and the length of speech in seconds in Study 2.}
    \label{tab:descriptive-unit2}
    \begin{tabular}{C{1.6cm}|c|c}
         & Word & Speech \\
        \hline
        Whole w/o Emma & 1655.7 (681.9) & 1520.1 (680.7)\\ \hline
        Student-student & 1303.6 (1413.0) & 1164.3 (581.5) \\
    \end{tabular}
\end{table}

\begin{table}[t]
    \centering
    \caption{The descriptive statistics of the balance of participation measures in Study 2.}
    \label{tab:descriptive-formula2}
    \begin{tabular}{C{1.6cm}|c|c|c|c}
         & & SD & IP & Entropy \\
        \hline
        \multirow{4}{1.6cm}{Whole w/o Emma} & \multirow{2}{*}{Word} & 174.2 & .130 & .021 \\
        & & (93.2) & (.113) & (.038) \\
        & \multirow{2}{*}{Speech} & 266.1 & .186 & .041 \\
        & & (230.7) & (.151) & (.048) \\ \hline
        \multirow{4}{1.3cm}{Student-student} & \multirow{2}{*}{Word} & 446.5 & .250 & .070 \\
        & & (859.1) & (.176) & (.116) \\
        & \multirow{2}{*}{Speech} & 226.1 & .214 & .057 \\
        & & (192.6) & (.181) & (.066) \\
    \end{tabular}
\end{table}

\begin{table}[t]
    \centering
    \caption{Partial correlation between participation balance and post-test scores controlled by pre-test scores in Study 2 ($n=13$). Correlations marked with * have $p<.05$ after being corrected with the Holm method.}
    \label{tab:results2}
    \begin{tabular}{C{1.6cm}|c|c|c|c}
         & & SD & IP & Entropy \\
        \hline
        \multirow{2}{1.6cm}{Whole w/o Emma} & Word & .488 & -.006 & -.209 \\
         & Speech & -.568 & -.619 & -.594 \\ \hline
        \multirow{2}{1.3cm}{Student-student} & Word & -.654 & -.890* & -.778* \\
        & Speech & -.464 & -.533 & -.451 \\
    \end{tabular}
\end{table}

\section{Discussion}
\subsection{Guidance on balance metrics}
We have shown how the learning outcomes of groups are connected with different metrics of participation balance in collaborative teaching to a teachable robot, Emma. We summarize the guidance on which metrics to use as a proxy of learning in collaborative learning with a robot.

\textbf{1. Balance in terms of the number of words can predict learning better than other units.} Study 1 revealed that only balance in terms of the number of words and the length of speech correlated with learning. Study 2 replicated this correlation with the number of words in automatically transcribed text but not with the speech length derived from audio waves. This could be because getting the correct speech length is challenging. You would need either annotation considering pauses within a turn or treatment of noise. In Study 2, although we carefully tuned the input volume of the microphones and applied a noise gate, the partners' and Emma's voices were still present because they sat close to each other (cf. Figure \ref{fig:emma}). However, the automatic transcription model (Whisper) labeled these segments as inaudible, which served as another layer of noise reduction. 

\textbf{2. CON, IP, and entropy may work better when the SD of the unit of participation is large compared to its mean.} In both studies, CON,\footnote{CON is equal to IP in Study 2, where the number of speakers is two in all portions.} IP, and entropy behaved similarly and showed significant correlations when SD and AD did not. One thing in common in the portions where CON, IP, and entropy had significant correlations was the SD of word counts was more than half of its mean. This was not the case in the whole dialogues excluding Emma's turns in Study 1. Future work can examine this hypothesis by preparing datasets that have small SDs of the units of participation.

\textbf{3. Balanced participation in the discussion between students is the best predictor of learning.} In Study 1, significant correlations with learning outcomes appeared only after removing Emma's turns from the analysis and originated from the student-student conversations. Study 2 saw significant correlations only in the student-student conversations.
When we looked at the two groups with the highest IP (i.e., most imbalanced) in Study 1, the students who spoke more scored lower in the post-test than in the pre-test. In those groups, when they had a long turn with explanations or thinking, their partners tended to confirm by simply saying, for example, ``Yeah'' or ``Okay.'' In more balanced groups, however, students were likely to elaborate on their partners' long turns. More work on understanding how this balance affects individual and collective learning in the presence of a teachable robot is needed.

\subsection{Limitations}
One limitation of this study is that many participants scored high on the pre-test and thus did not have much room for additional learning. The average pre-test scores were 75\% in Study 1 and 83\% in Study 2. Another limitation is both studies had a small number of samples ($n=13$ each). This small size prevented us from using demographic information as a potential moderator of an outcome. For instance, a group's heterogeneity in age and gender negatively affects participation balance in HRI \cite{skantze2017predicting}. This effect may exist in our studies, too, but, due to the small sample size, we focused on the relationship between participation balance and and groups' learning outcomes regardless of demographic diversity within a group. 

Second, our human-human-robot interaction may be different than other human-human-robot interactions in the asymmetrical nature of the interaction. Even though Emma participated in a discussion, students had to press a button to talk to her. A robot that participates in a richer way in the discussion such as automatically determining when to jump in without a button may yield different results.

Finally, we did not compare the metrics of balance among varying sizes of groups. We did not include CV or the Gini coefficient in our analysis because they were linear transformations of other formulas we used for a fixed group size. Moreover, turns might have not been a good unit just because they were fairly balanced in one-on-one student conversations. Future work can test these metrics with a dataset consisting of various sizes of groups.

\section{Conclusions}
This paper compares metrics of participation balance in terms of their ability to predict learning outcomes as a group in collaborative learning with a robot from three dimensions: units of participation, formulas for conversion, and portions of dialogue. The first study shows balance in word counts and speech length in student discussions correlates with learning. The second study confirms the correlations of balance in word counts only using the units of participation obtained from automated systems, simulating information available to a robot in a real-world scenario. Based on these studies, we provide guidance on which metrics of balance researchers should pick for scenarios similar to ours.


\section{Acknowledgments}
We would like to thank PETAL lab and FacetLab members at the University of Pittsburgh and anonymous reviewers for their thoughtful comments on this paper. This work was supported by Grant No. 2024645 from the National Science Foundation, Grant No. 220020483 from the James S. McDonnell Foundation, and a University of Pittsburgh Learning Research and Development Center internal award.

%
\bibliographystyle{abbrv}
\bibliography{sigproc}  
%

\balancecolumns
\end{document}